\documentclass[preprint,12pt,number,3p,twocolumn]{elsarticle}

\usepackage{amssymb}
\usepackage{lineno}
\usepackage{float}
\usepackage{graphics}

\usepackage[footnotesize]{caption2} 

\usepackage{threeparttable} %for tnote
\usepackage{multirow}
\usepackage{amsmath}%for equation
\usepackage{breqn}
\usepackage{flushend} %for last page 
\usepackage{color}
\usepackage{ulem} 

\journal{Nuclear Instruments and Methods in Physics Research Section A}

\begin{document}

\begin{frontmatter}



\title{Improvement of charge resolution for radioactive heavy ions at relativistic energies using a hybrid detector system}


\author[a]{J.W. Zhao}
\author[a,b,els]{B.H. Sun\corref{cor1}}
\ead{bhsun@buaa.edu.cn}
 %alpha-beta
\author[a]{L.C. He}
\author[a]{G.S. Li}
\author[a]{W.J. Lin}
\author[a]{C.Y. Liu}
\author[d]{Z. Liu}
\author[d]{C.G. Lu}
\author[a]{D.P. Shen}
\author[d]{Y.Z. Sun}
\author[d]{Z.Y. Sun}
\author[a,b]{I. Tanihata}
\author[a,b]{S. Terashima}
\author[e,f]{D.T. Tran}
\author[a]{F. Wang}
\author[a]{J. Wang}
\author[d]{S.T. Wang}
\author[a]{X.L. Wei}
\author[d]{X.D. Xu}
\author[a,b,els]{L.H. Zhu}
\author[a]{J.C. Zhang}
\author[d]{X.H. Zhang}  
\author[a]{Y. Zhang}
\author[a]{Z.T. Zhou}

\cortext[cor1]{Corresponding author} 
\address[a]{School of Physics and Nuclear Energy Engineering, Beihang University, Beijing, 100191, China}
\address[b]{International Research Center for Nuclei and Particles in Cosmos, Beihang University, Beijing, 100191, China}
\address[els]{Beijing Advanced Innovation Center for Big Data based Precision Medicine, Beihang University, 100083 Beijing, China}
\address[d]{Institute of Modern Physics, Chinese Academy of Sciences, Lanzhou 730000, China}
\address[e]{RCNP, Osaka University, Osaka 567-0047, Japan}
\address[f]{Institute of Physics, Vietnam Academy of Science and Technology, Hanoi 10000, Vietnam}
\iffalse
\fi

\begin{abstract}
	In typical nuclear physics experiments with radioactive ion beams (RIBs) selected by the in-flight separation technique, Si detectors or ionization chambers are usually equipped for the charge determination of RIBs. The obtained charge resolution relies on the performance of these detectors for energy loss determination, and this affects the particle identification capability of RIBs. We present an approach on improving the resolution of charge measurement for heavy ions by using the abundant energy loss information from different types of existing detectors along the beam line. Without altering the beam line and detectors, this approach can improve the charge resolution by more than 12\% relative to the multiple sampling ionization chamber of the best resolution. 
\end{abstract}


\end{frontmatter}

\section{Introduction}   
\label{}
In-flight fragment separator composed of dipole and quadrupole magnets, is one of the most important facilities for experimental nuclear physics in terms of transportation and unambiguous separation of radioactive ion beams (RIBs)~\cite{H.GeisselNIMB70(1992)286, T.KuboNIMB204(2003)97}. Such in-flight beam line is equipped with several different types of detectors for the purpose of identifying secondary radioactive beams. A commonly used particle identification (PID) method is known as the B$\rho$-TOF-$\Delta E$ technique. The detector system typically includes a  pair of thin plastic scintillators for Time-of-Flight (TOF) measurements, position-sensitive detectors for magnetic rigidity (B$\rho$) measurements and a multiple sampling ionization chamber (MUSIC) or a few Si detectors for energy loss ($\Delta E$) measurements. So far, the TOF determined by fast timing detectors can reach an excellent resolution down to a few tens of pico-seconds (e.g., ~\cite{J.ZhaoNIMA823(2016)41, W.LinCPC41(6)(2017)066001}). The position resolution of a few hundred micrometers allows a fairly good B$\rho$ determination (e.g., ~\cite{H.MiyaNIMB317(2013)701, Y.SunNIMA894(2018)72}, while the precision of $\Delta E$ measurements limit the charge resolution.

For heavy ions at relativistic energies, Si detectors and gas ionization chambers are the most commonly used $\Delta E$ detectors.
 A Si detector has substantially a good linear response over a large energy range and a good energy resolution due to the small average energy required to produce an ionization. However, its performance can be dramatically deteriorated with the radiation damages by heavy ions ~\cite{Si.(2018)}. On the other hand, a MUSIC has no radiation damage problem owing to the gas flow during operation. With the multiple sampling method, a fairly good energy resolution has been obtained ~\cite{K.KimuraNIMA538(2005)608, X.ZhangNIMA759(2015)389}. Moreover, it is easy to fabricate a large scale MUSIC. Plastic scintillators with typically a few milimeters thickness have also been used for charge determination in particular for low-$Z$ ions (e.g., in Ref.~\cite{P.MaarXiv}). Although its resolution is generally worse than that of a MUSIC or Si detector, its high counting-rate capability, very good noise immunity with photomultipliers as readout, good radiation hardness and simple construction procedures are the advantages for the charge determination. 

 From the perspective of statistical error, the larger the energy loss is, the more precise the charge determination is. In this paper, we aim to use all the abundant energy loss information from various types of detectors to improve the charge determination. As we will demonstrate, a consistent treatment of these information offers an efficient and economical way to improve the precision of $\Delta E$ measurement and hence results in a better charge resolution. 
 
The paper is organized as follows. Sec.~\ref{Sec2} presents the idea on how to use the energy loss information from different types of detectors to improve the charge determination. In Sec.~\ref{Sec3}, this approach is examined in a production experiment, and the results are then presented in Sec.~\ref{Sec4}. Finally, a summary is given in Sec.~\ref{Sec5}.

\section{Principle}
\label{Sec2}
\subsection{Methods to combine two detectors}

Let's start with two detectors and moreover neglect the possible electronics contribution to the energy loss determination. The measured $\Delta E$ follows a good Gaussian distribution, and its uncertainty can be described by $\sigma=\sqrt{F\Delta E\epsilon}$. $\epsilon$ is the average energy required to produce an ionization~\cite{W.Leo}, and the parameter $F$ links the energy loss and its fluctuation. $F$ and $\epsilon$ depend on the material of detectors, for example, $\epsilon$ for gas is about ten times larger than that of Si. 

For two detectors with energy deposits $\Delta E_{1}$  and  $\Delta E_{2}$, their uncertainties are
$\sigma_{1}=\sqrt{F_{1}\Delta E_{1}\epsilon_{1}}$ and $\sigma_{2}=\sqrt{F_{2}\Delta E_{2}\epsilon_{2}}$, respectively.
If $\Delta E_{1}=g^{2}\Delta E_{2}$ and $\sqrt{F_{1}\epsilon_{1}}=k\sqrt{F_{2}\epsilon_{2}}$, then $\sigma_{1}^2=k^2g^2\sigma_{2}^2$. $g^{2}$ is the coefficient to match energy deposits in two detectors, while $k$ compares the detector performance relevant to the energy resolution. The resulting resolution of two detectors can then be calculated as 

\begin{equation}
\frac{\sigma_{1}}{\Delta E_{1}}=\frac{k}{g}\frac{\sigma_{2}}{\Delta E_{2}}. 
\end{equation}

Different ways to combine energy deposits in such two detectors are depicted in details below. 

1) Arithmetic average: The $1^{st}$ method of energy loss combination is simply by $\Delta E=(\Delta E_{1}+\Delta E_{2})/2$, then the resulting resolution can be calculated as

\begin{equation}
\frac{\sigma}{\Delta E}=\frac{\sqrt{k^2g^2+1}}{g^2+1}\frac{\sigma_{2}}{\Delta E_{2}}. 
\end{equation}
where $\sigma=\sqrt{\sigma_{1}^2+\sigma_{2}^2}/2$.

2) Weighted average: The $2^{nd}$ method is the weighted average of $\Delta E_{1}$ and $\Delta E_{2}$. With weights of $w_1=1/\sigma_1^2$ and $w_2=1/\sigma_2^2$, $\Delta E=(w_1\Delta E_{1}+w_2\Delta E_{2})/(w_1+w_2)$ and its error $\sigma=\sqrt{1/(w_1+w_2)}$ can be obtained. Then the resolution is

\begin{equation}
\frac{\sigma}{\Delta E}=\frac{k\sqrt{k^{2}g^{2}+1}}{g(k^{2}+1)}\frac{\sigma_{2}}{\Delta E_{2}}. 
\end{equation}

3) Weighted average with the bias correction: 
The uncertainty $\sigma=\sqrt{F\Delta E\epsilon}$ is related to the energy deposit $\Delta E$ in the detector. The weights in the $2^{nd}$ method are biased due to the energy deposit difference in these two detectors. Then the performance of the $2^{nd}$ method will be deteriorated. To correct such bias, one can match the energy loss of one detector to that of another detector, i.e., using $\Delta E_{2}^{'}=g^{2}\Delta E_{2}=\Delta E_{1}$ ($\sigma_2^{'}=g^{2}\sigma_2$) to replace $\Delta E_{2}$ ($\sigma_2$) in the previous calculation. Nevertheless, the intrinsic resolution of the second detector remains the same, $\sigma_{2}^{'}/\Delta E_{2}^{'}=\sigma_{2}/\Delta E_{2}$.
To take the weighted average of $\Delta E_{1}$ and $\Delta E_{2}^{'}$, $\Delta E=(w_1\Delta E_{1}+w_2^{'}\Delta E_{2}{'})/(w_1+w_2^{'})$ with the weights $w_1=1/\sigma_{1}^{2}$, $w_{2}^{'}=1/\sigma_{2}^{'2}$ and $\sigma=\sqrt{1/(w_1+w_2^{'})}$ can be obtained. Then the resolution is 

\begin{equation}
\frac{\sigma}{\Delta E}=\frac{k}{\sqrt{k^{2}+g^{2}}}\frac{\sigma_{2}}{\Delta E_{2}}.
\end{equation}

\begin{figure}
	\centering
	\includegraphics[width=0.45\textwidth]{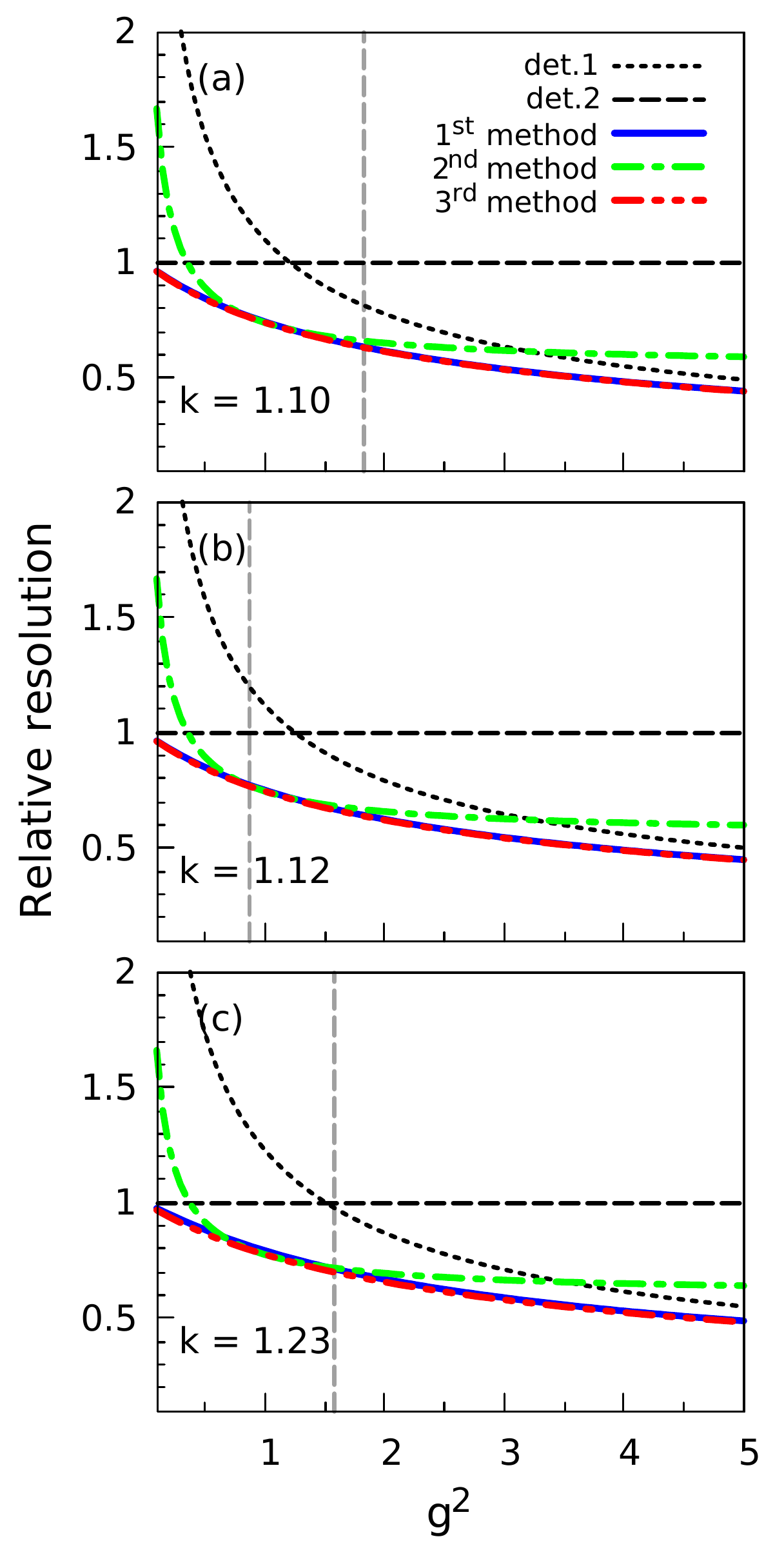}
	\caption{Resolutions normalized to det.2 in different methods for several selected values of $k$ and $g^{2}$ based on experimental data of TOF-start, MUSIC and Si detectors. (a) For TOF-start (det.1) and Si (det.2), k is around 1.10 and $g^{2}=1.82$. (b) For Si (det.1) and MUSIC (det.2), k is around 1.12 and $g^{2}=0.87$. (c) For TOF-start (det.1) and MUSIC (det.2), k is around 1.23 and $g^{2}=1.59$.  Values of $g^{2}$ in our experiment are indicated with vertical dashed lines, assuming a negligible contribution from the electronics and beam broadening.}
	\label{Fig1}
\end{figure}

Fig.~\ref{Fig1} shows the results of three different methods (labeled with `$1^{st}$ method', `$2^{nd}$ method' and `$3^{rd}$ method' accordingly) and that of the single detector (labeled with `det.1' and `det.2' in respectively). Several selected values of $k$ and $g^{2}$ are shown by referring to experimental data. The resolutions are normalized to the single detector `det.2'. 

The $3^{rd}$ method always offers the best improvement on resolution, and the improvement depends on the values of $k$ and $g^{2}$. The best performance of the $3^{rd}$ method is achieved when the resolution of these two detectors are same, i.e., $k/g=1$. The corresponding relative resolution is $1/\sqrt{2}$ in this case, i.e., $\sqrt{2}$ times better than the resolution of a single detector.

The $2^{nd}$ method leads to the same resolution as the $3^{rd}$ method when $g^{2}=1$. Its performance can be comparable with that of the $3^{rd}$ method only in a narrow range of $g^{2}\approx 1$, where no bias exists. When $g^{2}$ is far from 1, the degeneration in the resulting resolution becomes significant.  

The performance of the $1^{st}$ method will be exactly the same as that of the $3^{rd}$ method when $k=1$. 
This can be clearly verified by comparing Eq.~(2) with Eq.~(4), and can be seen in Fig.~\ref{Fig1} (a) and (b) for $k=1.10$ and 1.12. 
Even at $k=1.23$ as shown in Fig.~\ref{Fig1} (c), the performance of the $1^{st}$ method is only slightly worse than that of the $3^{rd}$ method. 
Therefore, the $1^{st}$ method can be used as a good approximation to the $3^{rd}$ method here.  

When $g^{2}$ is very far from 1, i.e., the energy deposit in one detector is much smaller than that in the second one, the first detector would not contribute much to the resulting resolution. As shown in Fig.~\ref{Fig1}, when $g^{2} \to 0$, the resolution obtained with the `$3^{rd}$ method' tends to that of the det.2. In contrast, when $g^{2}\gg1$, it is getting close to that of the det.1.
	
Similarly, when $k$ departs from 1, i.e., the intrinsic energy resolution of one detector is getting significantly worse than the other one, the first detector will not contribute much to the resulting resolution. Thus, the resolution is expected to be hardly improved by the combined method. This can be confirmed by examining Eq.~(2) and Eq.~(4). The resolution in the $3^{rd}$ method tends to that of the det.2 for $k\gg1$, and to that of the det.1 for $k\ll1$.
	 
As described above, the performance of the combination methods strongly depends on the values of $k$ and $g^{2}$. Thus, the optimum choice on the method for combining the energy loss relies on the knowledge of $k$ and $g^{2}$. An estimation of those values will be discussed based on experimental data in Sec. 4.

\subsection{Electronics contribution to the energy resolution}
In reality, the energy deposit in each detector is recorded with the following electronics and one can only obtain the digitalized energy loss value in channel number. Therefore, the ratio $g^{2}$ can be close to 1 by adjusting the gain of detectors and the following electronics in the experiment. For simplicity, we only discuss the case that energy losses for two detectors are the same here.

Assuming $\Delta E_{1}=\Delta E_{2}$ and $\sqrt{F_{1}\epsilon_{1}}=k\sqrt{F_{2}\epsilon_{2}}$, the resulting energy distribution can be calculated as:
\begin{equation}
\sigma_{1}=\sqrt{F_{1}\Delta E_{1}\epsilon_{1}+\sigma_{b}^{2}+\sigma_{e1}^{2}} \;,  
\end{equation}
and
\begin{equation} 
\sigma_{2}=\sqrt{F_{2}\Delta E_{2}\epsilon_{2}+\sigma_{b}^{2}+\sigma_{e2}^{2}} \;. 
\end{equation}

Here, $\sigma_{b}$ is the contribution from beam broadening, $\sigma_{e1}$ and $\sigma_{e2}$ are contributions from electronics (e.g., preamplifier, shaping amplifier and the analogy/charge to digital converters). By adopting same electronics and operation, $\sigma_{e1}$ and $\sigma_{e2}$ can be considered to be the same. Then we rewrite $\sigma_{b}^{2}+\sigma_{e1}^{2}=\sigma_{b}^{2}+\sigma_{e2}^{2}=\gamma F_{2}\Delta E_{2}\epsilon_{2}$. The resolution of the first detector can be rewritten as 

\begin{equation}
\frac{\sigma_{1}}{\Delta E_{1}}=\frac{\sqrt{k^{2}+\gamma}}{\sqrt{\gamma+1}}\frac{\sigma_{2}}{\Delta E_{2}}. 
\end{equation}

Accordingly, the resulting resolution in the $1^{st}$ method (arithmetic average) is

\begin{equation}
\frac{\sigma}{\Delta E}=\frac{\sqrt{k^{2}+2\gamma+1}}{2\sqrt{\gamma+1}}\frac{\sigma_{2}}{\Delta E_{2}}\;, 
\end{equation}
and the resolution with the $3^{nd}$ method (Weighted average) is 
\begin{equation}
\frac{\sigma}{\Delta E}=\frac{\sqrt{k^{2}+\gamma}}{\sqrt{k^{2}+2\gamma+1}}\frac{\sigma_{2}}{\Delta E_{2}}. 
\end{equation}

Since $\Delta E_{1}=\Delta E_{2}$, the $2^{rd}$ method is the same as the $3^{nd}$ method.

\begin{figure}[htb]
	\centering
	\includegraphics[width=0.45\textwidth]{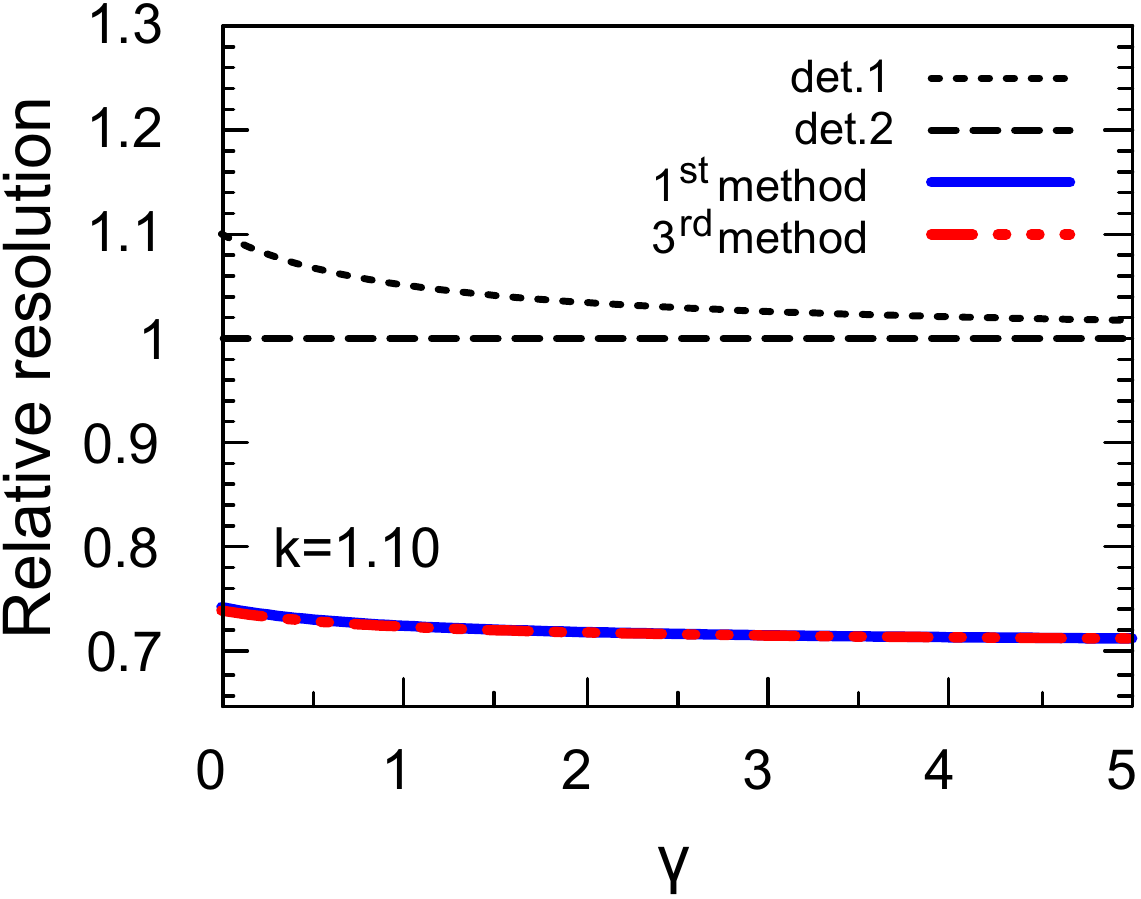}
	\caption{Relative resolution normalized to det.2 as a function of $\gamma$ at $k=1.10$.}
	\label{Fig2}
\end{figure}

The effect of contributions from electronics noise and beam broadening to the energy resolution is shown in Fig.~\ref{Fig2} at a fixed $k$ value of 1.10. Though the contribution from electronics noise and beam broadening is considered, there is no visual difference between the $1^{st}$ and the $3^{rd}$ methods. 
%\begin{enumerate}[(1)]
%	\item
%	\item 
%	\item 
%  \end{enumerate}  

The resolution of det.1 gets closer to that of det.2 with the increasing of $\gamma$, and eventually the resulting resolution $\sigma/\Delta E$ converges to $1/\sqrt{2}$ times of $\sigma_2/\Delta E_2$.

\section{Experiment}
\label{Sec3}
The experiment was performed at the fragment separator RIBLL2 at HIFRL-CSR in Lanzhou, China~\cite{J.XiaNIMA488(2002)11,X.ZhouNPN26(2016)4}. A primary beam of $^{18}$O at 400 MeV/nucleon impinged on the beryllium production target with a thickness of 30 mm. Secondary beams were produced via the projectile fragmentation reactions. Radioactive nuclei of interest were produced, separated in flight at a velocity of around 70\% speed of light with the first half of RIBLL2 and then delivered to the external target facility (ETF)~\cite{B.H.SunSciBul63(2018)78}. 

\begin{figure}[htb]
	\centering
	\includegraphics[width=0.45\textwidth]{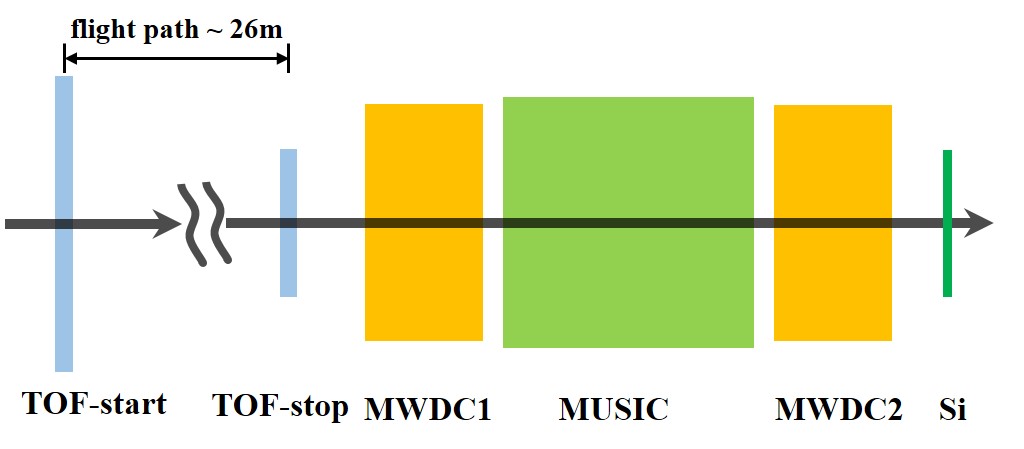}
	\caption{The experimental setup. TOF-start is installed at F1 of RIBLL2 and the other detectors are installed at ETF.}
	\label{Fig3}
\end{figure}

\begin{figure*}[htb]
	\centering
	\includegraphics[width=0.9\textwidth]{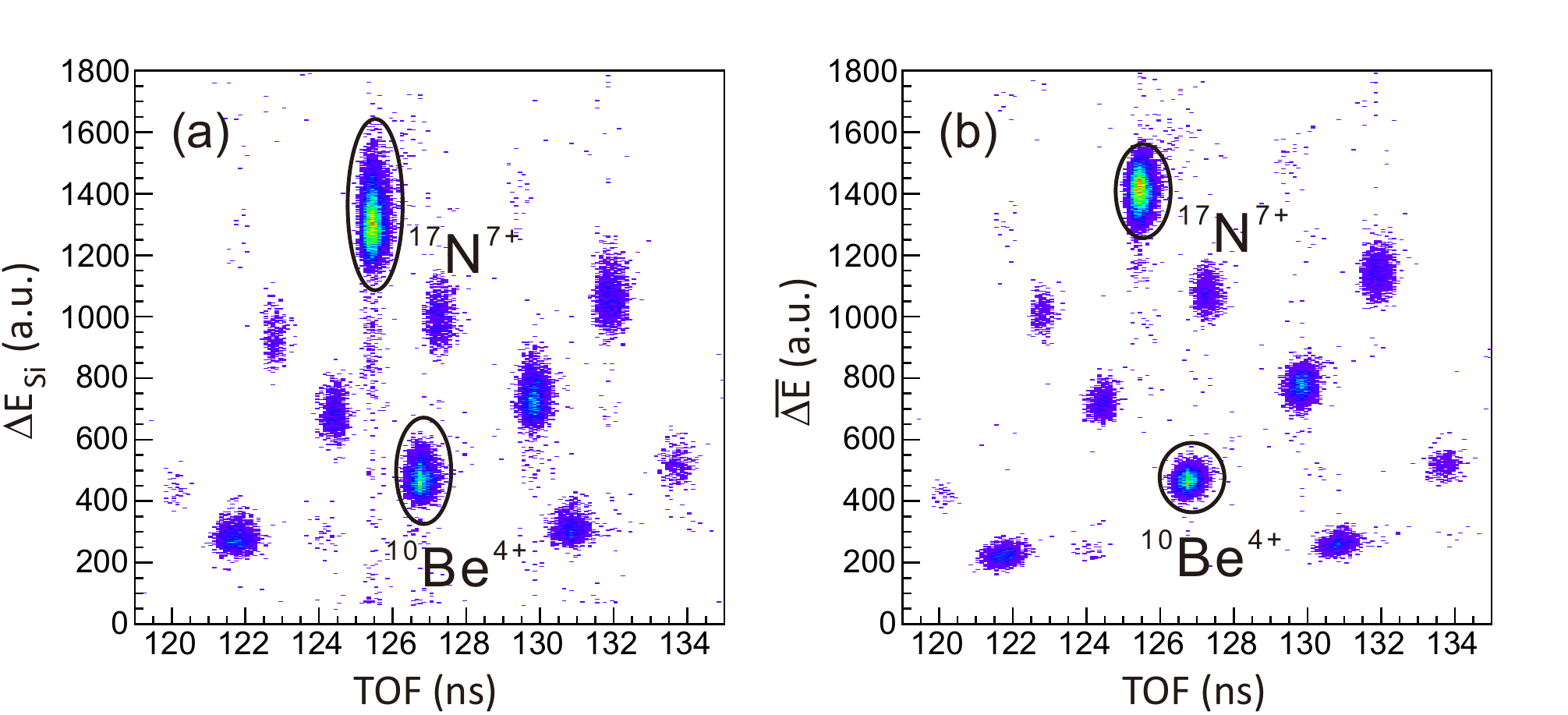}
	\caption{(a) `$\Delta E$ vs. TOF' plot obtained with the Si detector. (b) Same as (a) but obtained with the combined analysis of TOF-start, Si and MUSIC. The same ion species are much better grouped when using the combined energy loss information, $\overline{\Delta E}$. } 
	\label{Fig4}
\end{figure*}

The experiment setup is shown in Fig.~\ref{Fig3}. Two plastic scintillation counters (TOF-start and TOF-stop) installed at the foci F1 and ETF, were used to measure the TOF. The corresponding flight path from F1 to ETF is around 26 meters. A MUSIC~\cite{X.ZhangNIMA759(2015)389} and one Si detector were installed for charge determination. The track of each ion was determined by two multi-wire drift chambers (MWDC1 and MWDC2) placed upstream and downstream the MUSIC, respectively. Energy loss of the RIBs in the TOF-start, Si and MUSIC have been recorded. This makes it possible to examine different combinations of these abundant energy loss information for a better charge resolution. 

\section{Results and Discussion}
\label{Sec4}
The experimental `$\Delta E$ vs. TOF' plot with the energy loss obtained from the Si detector ($\Delta E_{Si}$) is shown in Fig.~\ref{Fig4} (a). To combine different detectors, the simplest way is the $1^{st}$ method discussed in Sec.~\ref{Sec2}. Then the arithmetic average energy loss of TOF-start, Si and MUSIC detectors in the experiment is 

\begin{equation}
\overline{\Delta E}=\frac{\Delta E_{TOF-start}+\Delta E_{Si}+\Delta E_{MUSIC}}{3} \;,
\end{equation}
where $\Delta E_{i}$ is the energy loss in detector $i$ recorded by the electronics such as ADC and QDC. Fig.~\ref{Fig4} (b) presents the `$\Delta E$ ̅vs. TOF' plot obtained with the arithmetic average energy loss $\overline{\Delta E}$. A much better isotopic
separation has been achieved by using $\overline{\Delta E}$ from the combined analysis. 
This is reflected by the narrow $\Delta E$ distribution for each isotopes and much less outliers originating from, e.g., the channeling effect in the silicon detector (see in particular the $^{17}$N$^{7+}$ case in Fig.~\ref{Fig4} (b)).

\begin{table*}
	\centering
	\setlength{\abovecaptionskip}{0pt}%    
	\setlength{\belowcaptionskip}{5pt}%
	\caption{Charge resolution $\sigma_{Z}$ of individual detector (TOF-start, MUSIC, Si) and results obtained with different methods of combining these three detectors.} % title name of the table
	\centering % centering table
	\begin{tabular}{l c c c c c c} % creating 7 columns
			\hline % inserting double-line
		Z & TOF-start & MUSIC & Si & $1^{st}$ method & $2^{nd}$ method & $3^{rd}$ method
		\\ [0.5ex]
		\hline % inserts single-line
		% Entering 1st row
		3 &  0.167 & 0.113  &  0.205 & 0.099 & 0.099 & 0.111 \\ [0.5ex]
		% Entering 2nd row
		4 &  0.158 & 0.112  &  0.194 & 0.093 & 0.092 & 0.100 \\ [0.5ex]
		% Entering 3rd row
		5 &  0.160 & 0.114 	&  0.180 & 0.091 & 0.089 & 0.100 \\ [0.5ex]
		% Entering 4th row
		6 &   0.169 & 0.122	&  0.194 & 0.091 & 0.087 & 0.098 \\ [0.5ex]	
		% Entering 5th row
		7 &  0.174 & 0.131 	&  0.199 & 0.097 & 0.095 & 0.106 \\ [0.1ex]		
	    % [0.5ex] adds vertical space
		\hline % inserts single-line
	\end{tabular}
%	\begin{tablenotes}
%		\item {*}
%	\end{tablenotes}
	\label{table1}
\end{table*}

\begin{figure}[htb]
	\centering
	\includegraphics[width=0.45\textwidth]{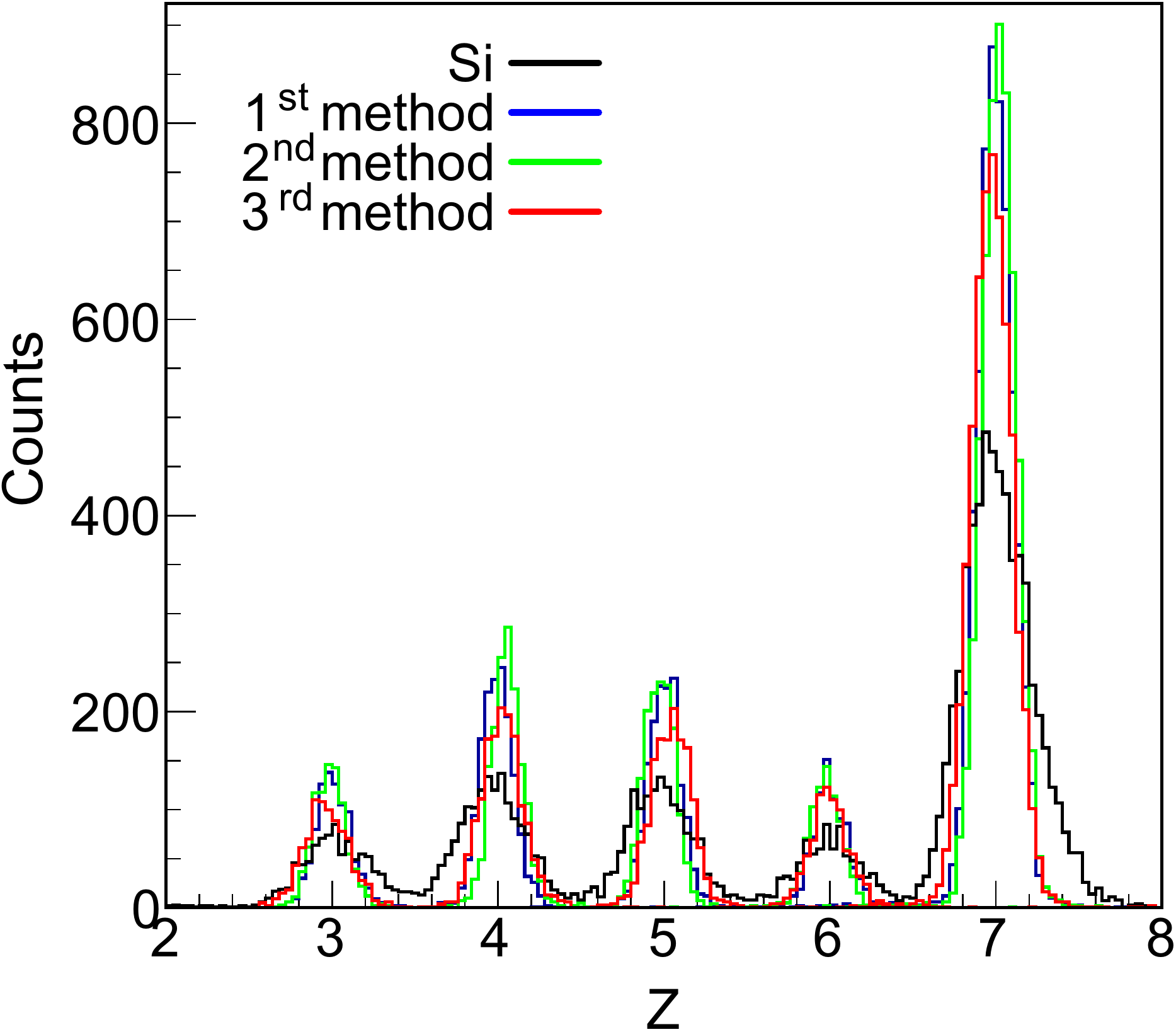}
	\caption{The charge distributions obtained with Si detector and the combined energy loss from plastic counter TOF-start, Si and MUSIC with the methods introduced in Sec.~\ref{Sec2}.}
	\label{Fig5}
\end{figure}

The charge distributions obtained with $\Delta E_{Si}$ and the combined analysis from plastic counter TOF-start, Si and MUSIC are depicted in Fig.~\ref{Fig5}. The $Z$ distribution is found to be well described by a Gaussian function. Therefore, the Gaussian parameter $\sigma_{Z}$ (the standard deviation of the corresponding $Z$ distribution) is adopted to evaluate the charge resolution $\sigma_{Z}$. The results are summarized as well in Table~\ref{table1}. The resulting charge resolution using the $1^{st}$ method is around 0.09. This is about 50\% better compared to the Si and 12\% to the MUSIC. Such an improvement is very useful to achieve a better PID capability for heavier isotopes. 

In Fig.~\ref{Fig4} (b), the  $1^{st}$  method is used. It is the most convenient and simplest way to combine different detectors. On the other hand, the computation of the $2^{nd}$ and $3^{rd}$ methods gets complicated because the resolution depends weakly on $Z$ as presented in Table~\ref{table1}.

Assuming negligible contribution from the electronics and beam broadening to the energy distribution in our experiment, the values of $k$ and $g^{2}$ can be estimated with experimental data. In this case,  
$g^{2}$ is the ratio of the digitalized energy loss value of one detector to that of the other one, 
and $k$ can be deduced from this $g^{2}$ and the digitalized energy loss distributions. 
The determined $g^{2}$ and $k$ values have been indicated with vertical dashed lines in Fig. 1.

In Fig.~\ref{Fig1} (a-c), when $k$ ranges from 1.10 to 1.23, the resolution in the $1^{st}$ method is comparable to that of the $3^{rd}$ method. Experimental data shows that three different approaches give similar resolution. The $3^{rd}$ method is even slightly worse. Practically, it is very hard to obtain the proper weights for the energy loss combination, because the contributions from the electronics and beam broadening of the secondary beam are not negligible and responses of detectors and electronics suffer nonlinearity. Then the $1^{st}$ method is considered to be the practically best choice to improve the charge resolution owing to the advantage on briefness over the $2^{nd}$ and $3^{rd}$ methods. 

\section{Summary}
\label{Sec5}
In this paper, we present a method to improve the charge resolution of RIBs by using the abundant energy loss information from existing detectors along the beam line. We formulated in details the approach by taking two detectors as an example. It is found that the charge resolution can be significantly improved by simply averaging the energy loss values obtained from two detectors. We verified this approach in an experiment and achieved an improvement of charge resolution by more than 12\% relative to the MUSIC, which has the best charge resolution. This approach is simple but useful, and can be applied to improve the charge resolution at the relativistic RIB experiment. 

\section*{Acknowledgments}
We appreciate the staff of the HIFRL-CSR accelerator for providing a stable beam during the experiment. This work was supported partially by the National Natural Science Foundation of China (Grant Nos. U1832211 and 11475014) and by the National Key R\&D program of China (2016YFA0400504).

\section*{References}

\biboptions{sort&compress}

\end{document}